\documentclass[prd,superscriptaddress,twocolumn,nofootinbib]{revtex4-2}
\usepackage{color}
\usepackage{graphicx}
\usepackage{amsfonts}
\usepackage{amssymb}
\usepackage{amsbsy}
\usepackage{amsmath}
\usepackage{mathrsfs}
\usepackage{latexsym}
\usepackage{natbib}
\usepackage{bm}
\usepackage{color}
\usepackage{graphicx}
\usepackage{hyperref}
\usepackage{float}
\definecolor{LinkColor}{rgb}{0.75, 0, 0}
\definecolor{CiteColor}{rgb}{0, 0.5, 0.5}
\definecolor{UrlColor}{rgb}{0, 0, 0.75}
\hypersetup{linkcolor=LinkColor}
\hypersetup{citecolor=CiteColor}
\hypersetup{urlcolor=UrlColor}
\begin{document}
\title{Local first law of black hole}
	
	        \author{Pabitra Tripathy}
	\email{pabitra.tripathy@saha.ac.in}
		\affiliation{
		Theory Division, Saha Institute of Nuclear Physics, 
		1/AF, Bidhannagar, Kolkata 700064, India}
	\affiliation{
		Homi Bhabha National Institute, Anushaktinagar, 
		Mumbai, Maharashtra 400094, India}
  
		\author{Pritam Nanda}
	\email{pritam.nanda@saha.ac.in}
	\affiliation{
		Theory Division, Saha Institute of Nuclear Physics, 
		1/AF, Bidhannagar, Kolkata 700064, India}
	\affiliation{
		Homi Bhabha National Institute, Anushaktinagar, 
		Mumbai, Maharashtra 400094, India}

  \author{Amit Ghosh}
	\email{amit.ghosh@saha.ac.in}
	\affiliation{
		Theory Division, Saha Institute of Nuclear Physics, 
		1/AF, Bidhannagar, Kolkata 700064, India}
	\affiliation{
		Homi Bhabha National Institute, Anushaktinagar, 
		Mumbai, Maharashtra 400094, India}

	\begin{abstract}
	We investigated the form and implications of the local first law of black hole thermodynamics in relation to an observer located at a finite distance from the black hole horizon. Our study is based on the quasilocal form of the first law for black hole thermodynamics, given by $\delta E=\frac{\bar{\kappa}}{8\pi}\delta A$, where $\delta E$ and $\delta A$ represent the changes in the black hole mass and area, respectively, and $\bar{\kappa}$ denotes the quasilocal surface gravity. We show that even at a finite distance, the quasilocal law still holds. It shows how the first law scales with the observer's location.
	\end{abstract}
	
	\maketitle 

 \section{Introduction}	 
 Hawking's semiclassical calculations \cite{Hawking:1975vcx} suggest that when a black hole is in a state of stationary equilibrium (reached after gravitational collapse), it behaves like a perfect black body by emitting thermal radiation, known as Hawking radiation, at a temperature proportional to its surface gravity. According to the first law of black hole mechanics, which relates changes in black hole's mass, area, and angular momentum, to its entropy (each black hole can be assigned an entropy). This entropy is proportional to the area of the black hole's event horizon. In other words, the surface area of a black hole is a measure of its entropy.\\ Finding a comprehensive statistical mechanical explanation for the thermal properties of black holes based on quantum theory remains a significant task for all proposed quantum theories of gravity. While attempts have been made to calculate the statistical entropy using string theory \cite{Strominger:1996sh} and loop quantum gravity \cite{Ashtekar:1997yu}, significant gaps in our understanding of the underlying quantum theories still exist in both approaches.\\
 An important challenge in addressing black holes in the framework of quantum gravity arises from the fact that the conventional definitions which rely on the global structure of spacetime. In the first law, while area, angular velocity, Coulomb potential, etc. are defined on the horizon, mass, angular momentum, charge, etc. are defined at asymptotic infinity. Furthermore, formation of the horizon itself requires knowledge about the full future of the spacetime. This issue has been recently highlighted in the context of two-dimensional models \cite{PhysRevLett.100.211302}. However, it is reasonable to expect that the physical concept of a large black hole emitting only a small amount of radiation and thus maintaining a state of near-equilibrium for an extended duration could be appropriately characterized by semiclassical physics. Such a characterization would be valuable for investigating the semiclassical regime of the underlying quantum theory in a more meaningful manner. Isolated horizons were proposed to offer a characterization of black holes in quasilocal manner. They capture the essential local properties of black hole event horizons while entirely dealing with quasilocal quantities only. Notably, isolated horizons adhere to a quasilocal variation of the first law, which describes the fundamental relationship between changes in the black hole's energy, area, and other physical quantities defined quasi-locally. 
 \begin{equation}\label{equ 1}
    \frac{\kappa_{IH}}{8\pi}\delta A=\delta\mathbb{E_{IH}}-\Omega_{IH} \delta\mathbb{L_{IH}}-\Phi_{IH}\delta\mathbb{Q_{IH}}
\end{equation}
Where $E_{IH}$, $L_{IH}$ and $Q_{IH}$ are the energy, angular momentum, and charge defined on the horizon and other quantities like $\kappa_{IH}$, $\Omega_{IH}$ and $\Phi_{IH}$ are already locally defined surface gravity, angular
velocity, and electrostatic potential on the isolated horizon. The aforementioned equation arises from the necessity for time evolution to adhere to the boundary conditions imposed by the isolated horizon (IH) and be Hamiltonian in nature \cite{PhysRevD.62.104025}. In an important work \cite{Frodden:2011eb}, the authors have pointed out that the first law of IH establishes that the isolated horizon energy $\mathbb{E_{IH}}$ must be a function of the energy $\mathbb{E_{IH}}(L_{IH},Q_{IH},A)$ of the system. The integrability conditions related to the previous phase space identity place limitations on the "intensive" properties. However, these conditions do not establish a favored concept of energy for the horizon according to the first law of IH. This limitation poses challenges for statistical mechanical explanations of quantum black holes. In their paper \cite{Frodden:2011eb} they also demonstrate that the aforementioned uncertainty is resolved when one comprehensively explores the quasilocal perspective, which was originally employed to define the isolated horizons. Interestingly, when examined by stationary observers positioned at an appropriate distance from the horizon, stationary black holes (and in a broader sense, isolated horizons) adhere to the quasilocal first law \cite{Frodden:2011eb}.
\begin{equation}\label{equ 2}
    \delta \mathsf{E}=\frac{\Tilde{\kappa}}{8\pi}\delta A
\end{equation}
Where $\Tilde{\kappa}=d^{-1}$, with $d^2<<A$, where $d$ is the proper distance of the observer from the black hole.\\

We observe that the laws of black hole mechanics depend significantly on observers. The first law of black hole mechanics involves quantities such as area, electric potential, angular velocity, etc. which are defined on the horizon, and also quantities such as ADM mass, electric charge, angular momentum, etc. at asymptotic infinity in an asymptotically flat spacetime. Thus, in order to write down the first law, we need to know the global structure of the spacetime. In this paper, we are attempting to write down a first law that is quasi-local. That is, the quantities needed to write down the first law are defined either on the horizon or at a finite proper distance from the horizon, and do not involve asymptotic structure of the spacetime. In an earlier work \cite{Frodden:2011eb}, the authors have already found such a first law in a spacetime region close to the horizon. From a practical standpoint, this issue appears to be more useful than a first law which requires the full knowledge of the global structure. This is particularly relevant in the case of astrophysical black holes as we ourselves are observers located at a finite proper distance from the black holes. In this paper, our objective is to address the issue of existence of this quasi-local version of the first law, and we find that the answer is indeed affirmative. Considering an observer located at a finite proper distance from the black hole, we can derive a first law, similar to equation \ref{equ 2}. However, in this case, the proper distance not being less than the area of the black hole surface, the quantity $\Tilde{k}$ becomes a complicated function of proper distance $(d)$.\\

The plan of the paper is outlined as follows:

In Section \ref{definition}, we provide a concise review of the thought experiment that establishes the quasilocal form of the first law. Additionally, we discuss the interpretation of effective surface gravity and temperature for the nearest observer in this context.

In Section \ref{finite}, we validate the conjecture's usefulness for finite observers. We explicitly analyze the scenario for RN black hole, Kerr black hole, and BTZ black hole, and conclude with a discussion on the significance of this model.

\section{Review of the quasilocal first law of black hole thermodynamics}
\label{definition}
\subsection{A gedanken experiment}
In this section, we will review a thought experiment conducted by Frodden et al\cite{Frodden:2011eb}.

First, let's consider a family of stationary observers located at a distance $\textit{d}$ from the stationary black hole spacetime. In a generic stationary black hole spacetime, there exist two killing vector fields: $t^\alpha$ and $\phi^\alpha$. These vector fields correspond to time translation and axial symmetry of the spacetime respectively. The family of stationary observers, denoted as [$\mathcal{P}$] follows the orbits determined by the killing vector fields  $\xi^\alpha$. This means at each point along the orbital path, the tangent vectors to the path align with $\xi^\alpha$. In the case of a stationary spacetime, these vector fields are given by the following expressions,
\begin{equation}
\label{obs}
 \xi^\alpha=t^\alpha+\Omega_h \phi^\alpha
\end{equation}
The angular velocity at the horizon of the black hole is denoted by $\Omega_h$. The stationary observers that are in corotation with the blackhole have an angular velocity equal to $\Omega_h$, which is given by,
\begin{equation}
    \Omega_h=\frac{a}{r_+^2+a^2}
\end{equation}

Here, $a$ represents the ratio of angular momentum ($J$) to the mass ($M$) of the stationary black hole, expressed as $J=aM$. The symbol $r_+$ corresponds to the event horizon of the black hole.
\par The four-velocity of the observers $\mathcal{P}$ is given by the normalized tangent vector field along their worldline. 
 \begin{equation}
 \label{normalized}
     \gamma^\alpha=\frac{\xi^\alpha}{\rVert {\xi^\alpha}\rVert}
 \end{equation}
The selection of these particular observers (\ref{obs}) is crucial for the argument presented in this study, as they possess the symmetries of the spacetime. The symmetries exhibited by these observers play a fundamental role in compactifying the first law. 
\par Let us examine a scenario in which a charged particle with unit mass and charge $e$ approaches the black hole from infinity and becomes absorbed by it. To account for general cases, we consider a background spacetime that is both charged and rotating. If the particle is moving with four-velocity $\eta^\alpha$, then its conserved energy and angular momentum are given by the following expressions, respectively:
\begin{equation}
\label{E}
    \mathbb{E}=-\eta^\alpha t_\alpha-e\mathcal{A}^\alpha t_\alpha
\end{equation}
\begin{equation}
\label{L}
    \mathbb{L}=\eta^\alpha\phi_\alpha+e\mathcal{A}^\alpha\phi_\alpha
\end{equation}
Where $\mathcal{A}_\alpha$ represents the electromagnetic four-potential resulting from the charge of the black hole.
\par The first law of black hole thermodynamics is an analogue of the first law of thermodynamics, applied specifically to black holes. It states that the change in energy of a black hole is related to the change in its mass, angular momentum, and electric charge, as well as the energy of the matter falling into the black hole and the energy carried away by emitted radiation.
Mathematically, the first law of black hole thermodynamics can be expressed as \cite{Bardeen:1973gs}:
\begin{equation}
\label{firstlaw}
    \delta M=\frac{\kappa}{8\pi}\delta A+\Omega_h \delta J+\Phi \delta Q
\end{equation}
Where $\delta M$ represents the change in the black hole mass, $\delta A$ represents the change in its horizon area, $\kappa$ denotes the surface gravity of the black hole, $\Omega_h$ denotes its angular velocity, $\delta J$ represents the change in its angular momentum, $\Phi$ represents its electric potential, and $\delta Q$ represents the change in its electric charge.
\par Now, in the context of the black hole absorbing the charged particle, the changes in the black hole's parameters will be linked to the charge, angular momentum, and energy of the particle. Specifically, we have the following relationships:
\begin{equation}
\label{change}
    \delta J=\mathbb{L}; \;\; \delta Q= e; \;\;  \delta M=\mathbb{E}
\end{equation}
By employing equation (\ref{change}), equation (\ref{firstlaw}) can be reformulated as follows:
\begin{equation}
\label{1stl}
    \frac{\kappa}{8\pi}\delta A=\mathbb{E}-\Omega_h \mathbb{L}-e\Phi
\end{equation}
 Now the local energy of the particle as measured by the observer $\mathcal{P}$ is given by,
\begin{equation}
\label{locale}
    \mathbb{E}_l=-\eta^\alpha \gamma_\alpha
\end{equation}
Using (\ref{obs}) and (\ref{normalized}) we can rewrite equation (\ref{locale}) as follows,
\begin{equation}
    \mathbb{E}_l=-\frac{\eta^\alpha t_\alpha+\Omega_h \eta^\alpha \phi_\alpha}{\rVert {\xi^\alpha}\rVert}
\end{equation}
Now, we can utilize equations (\ref{E}) and (\ref{L}) to express the local energy of the particle in terms of its conserved energy and angular momentum as follows
\begin{equation}
    \mathbb{E}_l=\frac{\mathbb{E}-\Omega_h\mathbb{L}+e\mathcal{A}^\alpha \xi_\alpha}{\rVert {\xi^\alpha}\rVert}
\end{equation}
The electric potential $\Phi$ at the horizon, as described in \cite{Wald:1984rg}, can be expressed in terms of the electromagnetic four-potential $\mathcal{A}_\alpha$ and the killing vector field $\xi^\alpha$ as follows:
\begin{equation}
\label{Phi}
     \Phi=-\mathcal{A}^\alpha \xi_\alpha
\end{equation}
By incorporating this definition of $\Phi$, we obtain a comprehensive expression for $\mathbb{E}_l$ as follows:
\begin{equation}
\label{El}
    \mathbb{E}_l=\frac{\mathbb{E}-\Omega_h\mathbb{L}-e\Phi}{\rVert {\xi^\alpha}\rVert} 
\end{equation}
Finally, by utilizing equations (\ref{1stl}) and (\ref{El}), we can establish a relationship between the local energy of the charged particle and the area of the black hole, up to a proportionality factor. Subsequently, we have:
\begin{equation}
\label{firstl}
    \mathbb{E}_l=\frac{\Tilde{\kappa}}{8\pi}\delta A
\end{equation}
Where,
\begin{equation}
\label{kbar}
    \Tilde{\kappa}=\frac{\kappa}{\rVert {\xi^\alpha}\rVert}
\end{equation}
It is important to note that from the perspective of the observer, the amount of energy absorbed by the black hole is given by $\mathbb{E}_l$, which must be equal to the increase of BH energy. For an observer who follows the integral curves of the killing vector field of the spacetime, the form of equation (\ref{firstlaw}) is simplified to,
\begin{equation}
\label{quasi1st}
    \delta \mathsf{E}=\frac{\Tilde{\kappa}}{8\pi}\delta A
\end{equation}

\subsection{Quasilocalness}
It is crucial to note that thus far, we have not imposed any restrictions on our stationary observer. As long as they follow the integral curves of the killing vector fields of the stationary black hole, equation (\ref{quasi1st}) remains well-defined. It is important to emphasize that $\Tilde{\kappa}$ is no longer the surface gravity; rather, it represents the ratio of the surface gravity to the norm of the killing vector field. At this point, an essential question arises: can we consider expression (\ref{quasi1st}) as the first law? To address this question in this section, we impose a restriction on our observers, namely that their distance $d$ from the black hole is very small, satisfying $d^2 << A$.
\par By performing a straightforward calculation (for explicit calculations, please refer to the next section), using the Kerr-Newman spacetime, it can be shown that:
\begin{equation}
\label{barkappa}
    \Tilde{\kappa}\approx \frac{1}{d}
\end{equation}
If we incorporate this expression of $\Tilde{\kappa}$ into equation (\ref{quasi1st}) then,
\begin{equation}
\label{final1st}
     \delta \mathsf{E}=\frac{1}{8\pi d}\delta A
\end{equation}
This expression can be interpreted as the quasilocal representation of the first law. The term $\Tilde{\kappa}$ is referred to as the quasilocal surface gravity, which remains independent of the mass, charge, and angular momentum of the black hole. This quasilocal first law establishes a unique relationship between the variation in energy and the black hole area for any stationary spacetime. It is noteworthy that the quasilocal version of the first law is universal, meaning it does not depend on whether the black hole is charged or neutral, rotating or static.
\subsection{Local temperature}\label{LT}
In this section, we will attempt to reinforce our argument that equation (\ref{quasi1st}) is indeed a valid equation of the local first law of black hole thermodynamics. To support our assertion, we must demonstrate that $\frac{\Tilde{\kappa}}{2\pi}$ is equivalent to the local temperature. To begin, we will define the local temperature using the methodology introduced by Frodden, Ghosh, and Perez\cite{frodden2011local}. Subsequently, we will express the Hawking flux in terms of this newly defined local temperature.
\par The local frequency $\omega_{loc}$ of a particle with a wave four-vector $\mathbf{k}_\alpha$, as measured by observer $\mathcal{P}$, can be expressed as:
\begin{equation}
    \label{lfreqn}
    \omega_{loc}=\mathbf{k_\alpha}\gamma^\alpha=\frac{\mathbf{k}_\alpha t^\alpha+\Omega_h \mathbf{k}_\alpha\phi^\alpha}{\rVert \xi^\alpha\lVert}
\end{equation}
Where $\gamma^\alpha$ represents the four-velocity of observers, as given by equation (\ref{normalized}).
Considering that the metric and electromagnetic fields are both time-independent and axially symmetric, it implies that $\xi^\alpha$, the Killing vector, Lie drags both the metric and electromagnetic field. As a consequence, two constants of motion arise:
\begin{equation}
    \begin{aligned}
        \omega&=\mathbf{k}_\alpha t^\alpha+e\mathcal{A}_\alpha t^\alpha\\
        j&=\mathbf{k}_\alpha \phi^\alpha+e\mathcal{A}_\alpha \phi^\alpha
    \end{aligned}
\end{equation}
Here, $\omega$ and $j$ represent the frequency and angular momentum of the particle measured at asymptotic infinity.
By substituting the expressions of $\omega,j$ and equation (\ref{Phi}), we can rewrite $\omega_{loc}$ as follows:
\begin{equation}
\label{omegal}
    \omega_{loc}=\frac{\omega-\Omega_h j-e\Phi}{\lVert \xi^\alpha\rVert}
\end{equation}
In contrast to the conventional Hawking calculation, the presence of rotation and charge in the Kerr-Newmann black hole leads to a shift in the frequency($\omega$) within the expression of Hawking flux, resulting in ($\omega-\Omega_hj-e\Phi$)\cite{Hawking:1975vcx}. Then one gets the number of particles emitted in the form,
\begin{equation}
    <\mathcal{N}>=\frac{\Gamma}{e^{2\pi \kappa^{-1}(\omega-\Omega_h j-e\Phi)}-1}
\end{equation}
Coefficient $\Gamma$ is called the grey body factor.
\par By utilizing equation (\ref{omegal}) the previous expression can be rewritten in terms of local frequency as follows,
\begin{equation}
     <\mathcal{N}>=\frac{\Gamma}{e^{\frac{2\pi\lVert\xi^\alpha\rVert}{\kappa}\omega_{loc}}-1}
\end{equation}
Which effectively captures the Planckian spectrum. With the help of equation (\ref{kbar}) we can express it as follows,
\begin{equation}
\label{HawSpectum}
    <\mathcal{N}_{loc}>=\frac{\Gamma_{loc}}{e^{\frac{2\pi}{\Tilde{\kappa}}\omega_{loc}}-1}
\end{equation}
This equation suggests that local observers perceive the aforementioned spectrum, and $\Tilde{\kappa}/2\pi$ can be understood as the local temperature. Here we have assumed that the nature of the thermal spectrum will preserve its global structure.

\section{Extension from quasilocal to finite observer}
\label{finite}
In the previous section, we derived the universal form of the first law for quasilocal observers under the assumption that the distance $d$ is very small. Now, we want to explore the possibility of observers staying at a finite distance from the black hole. As we mentioned earlier, the derivation of equation (\ref{quasi1st}) is independent of any specific choice of observer position. Furthermore, we have observed that the quasilocal surface gravity is inversely proportional to the distance $d$ (as shown in equation \ref{final1st}). Now, the question arises: can equation (\ref{barkappa}) be expressed in terms of a constant parameter for a finite distant observer? In this section, we will extend our investigation to address this question explicitly, focusing on the Reissner-Nordstrom, Kerr, and BTZ black holes.
\subsection{Reissner-Nordstrom BH}
In the ingoing Eddington–Finkelstein coordinates, the Reissner-Nordstrom (RN) solution is expressed as follows:
\begin{equation}
    ds^2=-f(r)dv^2+2dv dr+r^2 d\Omega^2
\end{equation}
Where $f(r)=\left(1-\frac{2M}{r}+\frac{Q^2}{r^2}\right)$. This function has zeros at $r=r_\pm$, where $r=r_+=M+\sqrt{M^2-Q^2}$ and $r=r_-=M-\sqrt{M^2-Q^2}$ are referred to as the inner and outer horizons, respectively.
\par The killing vector field of this spacetime is $\xi^\alpha=\partial_v^\alpha$. Our observers $\mathcal{P}$ follow the integral curves of $\xi^\alpha$. The norm of $\xi^\alpha$ is 
\begin{equation}
\label{zi}
    \lVert \xi \rVert=\sqrt{f(r)}=\sqrt{\frac{(r-r_+)(r-r_-)}{r^2}}
\end{equation}
The surface gravity of this spacetime is given by,
\begin{equation}
\label{kappa}
    \kappa=\frac{r_+-r_-}{2r_+^2}
\end{equation}
Now, we aim to measure the proper distance $d$ from the black hole to the observer along a curve $\mathcal{C}$ that is normal to both the event horizon and the orbits of the observers. Let us consider the tangent vector field to $\mathcal{C}$ as $K^\alpha = m(r)\partial_v^\alpha + n(r)\partial_r^\alpha$, where $m(r)$ and $n(r)$ are two arbitrary functions of $r$. It is important to note that as $\mathcal{C}$ is normal to the horizon, $K^\alpha$ does not have any $\theta$ or $\phi$ components (the horizon is foliated by topological two spheres). Furthermore, $K^\alpha$ is also normal to $\xi^\alpha$, i.e., $K^\alpha \xi_\alpha = 0$. This condition provides a relationship between $m$ and $n$, specifically $n(r) = f(r) m(r)$, where $f(r)$ is the function defined earlier. By using this relation, the tangent vector field to $\mathcal{C}$ can be expressed as $K^\alpha = m(r)[\partial_v^\alpha + f(r)\partial_r^\alpha]$. The value of $m(r)$ can be determined by the normalization condition, i.e., $K^\alpha K_\alpha = 1$. This gives $m(r) = \frac{1}{\sqrt{f(r)}}$. Finally, the tangent vector field to the curve $\mathcal{C}$ can be written as:
\begin{equation}
K^\alpha=\frac{1}{\sqrt{f(r)}}(\partial_v^\alpha+f(r)\partial_r^\alpha)
\end{equation}
Once we have defined our curve, it becomes straightforward to determine the proper distance from the black hole to the observer along the curve $\mathcal{C}$. This proper distance is given by:
\begin{align}
    d=\int_{r_+}^r\sqrt{g_{\alpha \beta} K^\alpha K^\beta}d\lambda
\end{align}
Where $\lambda$ is the parameter along the curve. Using the relation $K^r=\frac{dr}{d\lambda}=\frac{1}{\sqrt{f(r)}}$, we can rewrite the above equation as follows,
\begin{equation}
\begin{aligned}
\label{dist1}
       d&=\int_{r_+}^r\sqrt{g_{\alpha \beta} K^\alpha K^\beta}\frac{dr^\prime}{\sqrt{f(r^\prime)}} \\
       &=\frac{1}{2}(r_++r_-)\ln{(2r-r_+-r_-+2\sqrt{(r-r_+)(r-r_-)})}\\&+\sqrt{(r-r_+)(r-r_-)}-\frac{1}{2}(r_++r_-)\ln{(r_+-r_-)}
\end{aligned}
\end{equation}
Our goal is to express $\lVert \xi \rVert$ in terms of $d$. To achieve this, we need to solve equation (\ref{dist1}) for $r$ and substitute $r$ into equation (\ref{zi}). However, solving these complicated nonlinear equations analytically can be challenging. Therefore, without loss of generality, we can make an approximation that when the observer is at a large distance, $\sqrt{r-r_+} \approx \sqrt{r-r_-}$. We only employ this approximation to simplify the square root within the logarithmic term. By using this approximation, the expression for $d$ can be reformulated as:
\begin{equation}
\label{l1}
    d=\frac{(r_++r_-)}{2}(\ln{(4r-3r_+-r_-)}-\ln{(r_+-r_-)})+(r-r_+)
\end{equation}
%Where $l=d+\frac{1}{2}(r_++r_-)\ln{(r_+-r_-)}$ represents the distance scaled by a constant.
We observe that equation (\ref{l1}) is a transcendental equation that can be solved using the LambertW($\mathcal{W}$) function \cite{article}. Substituting the solution of equation (\ref{l1}) into equation (\ref{dist1}), we obtain the relation for $r$ in terms of $d$ as follows:
\begin{equation}
\label{r}
\begin{split}
r(d)=\frac{(r_++r_-)}{2}{\mathcal{W}\left(\frac{e^\frac{2(r_++r_-)\ln{(r_+-r_-)}+r_+-r_-+4d}{2(r_++r_-)}}{2(r_++r_-)}\right)}\\+{\frac{1}{4}(3r_++r_-)}
\end{split}
\end{equation}
By substituting the value of $r$ obtained from equation (\ref{r}), and utilizing the aforementioned approximation, equation (\ref{zi}) can be expressed as follows:
\begin{equation}
\label{zi1}
\begin{split}
    &\lVert \xi \rVert=1-\\ &\small\frac{r_+}{\frac{(r_++r_-)}{2}{\mathcal{W}\left(\frac{e^\frac{2(r_++r_-)\ln{(r_+-r_-)}+r_+-r_-+4d}{2(r_++r_-)}}{2(r_++r_-)}\right)}+{\frac{1}{4}(3r_++r_-)}}
\end{split}
\end{equation}
By utilizing equations (\ref{quasi1st}), (\ref{kappa}), and (\ref{zi1}), we can express the first law for a Reissner-Nordstrom black hole in terms of energy, area, and the proper distance from the black hole to the observer as follows:
\begin{equation}
\begin{aligned}
\label{rn}
\delta \mathsf{E}&=\frac{r_+-r_-}{16\pi r_+^2}\times \\&\frac{1}{\left(1-\frac{r_+}{\frac{(r_++r_-)}{2}{\mathcal{W}\left(\frac{e^\frac{2(r_++r_-)\ln{(r_+-r_-)}+r_+-r_-+4d}{2(r_++r_-)}}{2(r_++r_-)}\right)}+{\frac{1}{4}(3r_++r_-)}}\right)}\delta A\\
    &=\frac{\zeta_{RN}(d)}{8\pi}\delta A
\end{aligned}
\end{equation}

Where, $\zeta_{RN}(d)=\\\frac{r_+-r_-}{2 r_+^2\left(1-\frac{r_+}{\frac{(r_++r_-)}{2}{\mathcal{W}\left(\frac{e^\frac{2(r_++r_-)\ln{(r_+-r_-)}+r_+-r_-+4d}{2(r_++r_-)}}{2(r_++r_-)}\right)}+{\frac{1}{4}(3r_++r_-)}}\right)}$ is a function of proper distance $d$.
\par Equation (\ref{rn}) presents a straightforward first law of black hole mechanics for the Reissner-Nordstrom spacetime. It reveals a clear relationship between the change in the black hole's energy and its area. This relationship highlights the dependence on the observer through their proper distance from the black hole. In the quasilocal case, the same principle holds true, and the proportionality factor between the change in energy and the change in area is inversely proportional to the proper distance. However, in this case, the proportionality factor is expressed through a more complex function of $d$.\\

\begin{figure}[ht!]
     \includegraphics[width=7cm]{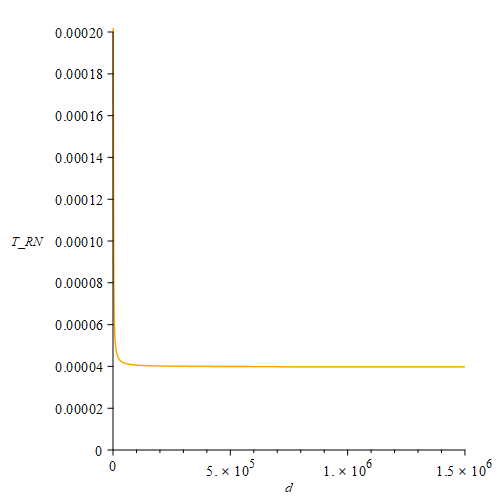}
     \caption{Here, we have plotted $T_{{loc}}$ as a function of the proper distance $d$ for the Reissner-Nordström (RN) black hole. For this plot, we have chosen the parameters M=1000 and Q=1. With these parameter values, the corresponding Hawking temperature is $T_{\text{H}}=3.97\times10^{-5}$. We have taken G=c=$k_B$=1.}
     \label{fig:RN}
 \end{figure}

As discussed in Section \ref{LT}, we observed that the Hawking spectrum maintains its form  (equn \ref{HawSpectum}) with a locally defined grey body factor and local energy. This allows us to introduce the concept of a local temperature. So In a Reissner-Nordström (R-N) spacetime, if an observer is situated at a finite distance away from the black hole, they will perceive the temperature as
\begin{equation}
    T_{loc}=\frac{\Tilde{\kappa}}{2\pi}=\frac{\zeta_{RN}(d)}{2\pi}
\end{equation}
 we can observe that the local temperature is solely dependent on the proper distance between the observer and the black hole. In other words, the temperature experienced by the observer is determined by how far they are from the black hole, without any additional factors influencing it.\\
 We have plotted $T_{\text{loc}}$ as a function of proper distance [see \ref{fig:RN}]. From the plot, it is evident that as the proper distance increases, the local temperature gradually approaches and merges with the Hawking temperature.
 
\subsection{ Kerr BH}
In this section, we will study the case of a rotating black hole. The Kerr metric in standard Boyer-Lindquist coordinates is given by \cite{Poisson:2009pwt} 
\begin{equation}
\begin{split}
    ds^2=-\left(1-\frac{2Mr}{\rho^2}\right)&dt^2-\frac{4Mar\sin^2{\theta}}{\rho^2}dtd\phi\\
    &+\frac{\Sigma}{\rho^2}\sin^2{\theta}d\phi^2+\frac{\rho^2}{\Delta}dr^2+\rho^2d\theta^2
    \end{split}
\end{equation}
Where
\begin{equation}
    \begin{split}
    &\rho^2=r^2+a^2\cos{\theta}^2\\
    &\Delta=r^2-2Mr+a^2\\
    &\Sigma=(r^2+a^2)^2-a^2\Delta\sin^2{\theta}
    \end{split}
\end{equation}
The spacetime exhibits a coordinate singularity at $\Delta=0$, which corresponds to the horizon of the Kerr black hole. The equation $\Delta=(r-r_+)(r-r_-)=0$ determines the locations of the two horizons, where $r=r_+=M+\sqrt{M^2-a^2}$ corresponds to the outer horizon and $r=r_-=M-\sqrt{M^2-a^2}$ corresponds to the inner horizon.
\par Due to the stationarity and axisymmetric nature of the spacetime, it possesses two killing vector fields, one corresponding to time translation and the other to rotational symmetry. The Observer, $\mathcal{P}$ follows the integral curves of the killing vector field of the spacetime, which is given by,  
\begin{equation}
    \xi^\alpha=t^\alpha+\Omega_h \phi^\alpha
\end{equation}
Where $\Omega_h=\frac{a}{r_+^2+a^2}$.The norm of the $\xi^\alpha$ vector field can be calculated as follows,
\begin{equation}
\label{zi2}
    \lVert \xi \rVert=\sqrt{1-\frac{2Mr}{\rho^2}+\Omega_h\left(-\frac{\Sigma}{\rho^2}\sin^2{\theta}\Omega_h+\frac{4Mar\sin^2{\theta}}{\rho^2}\right)}
\end{equation}
The surface gravity of a black hole can be calculated as follows,
\begin{equation}
\label{kappa2}
    \kappa=\frac{r_+-M}{r_+^2+a^2}
\end{equation}
To calculate the proper distance $d$ along the radial path $\mathcal{C}$, we can integrate the norm of the tangent vector $J^\alpha=\partial_r^\alpha$ over the path. The proper distance $d$ can be expressed as:
\begin{equation}
    \begin{aligned}
      d=\int_{r_+}^r\sqrt{g_{\alpha \beta} J^\alpha J^\beta}d\lambda
    \end{aligned}
\end{equation}
By normalizing the vector field $J^\alpha$, we can rewrite the above equation as follows:
\begin{equation}
    \begin{aligned}
      d=\int_{r_+}^r\frac{\rho}{\sqrt{\Delta}}dr^\prime
%    &= \int_{r_+}^r\frac{r^\prime}{\sqrt{(r^\prime-r_+)(r^\prime-r_-)}}dr^\prime\\
\end{aligned}
\end{equation}
By fixing the surface at $\theta = \frac{\pi}{2}$ and using the approximation $\sqrt{r-r_+} \approx \sqrt{r-r_-}$, the proper distance can be expressed as:
 \begin{equation}
    d=\frac{(r_++r_-)}{2}(\ln{(4r-3r_+-r_-)}-\ln{(r_+-r_-)})+(r-r_+)
\end{equation}
Using the expression for $r$ in terms of $d$ obtained earlier, we can rewrite equation (\ref{zi2}) in terms of $d$. Substituting this expression, as well as the value of $\kappa$ from equation (\ref{kappa2}), into equation (\ref{quasi1st}), we can obtain the explicit form of the first law for the Kerr metric for a finite distance observer.
\par The explicit form of the first law for the Kerr metric, taking into account a finite distance observer, can be written as follows:
\begin{equation}
\label{KNFL}
    \delta \mathsf{E}=\frac{\zeta_{KERR}(d)}{8\pi}\delta A
\end{equation}
Where,\\ $\zeta_{KERR}=\frac{r_+-M}{r_+^2+a^2}\times\\ \frac{1}{{\left(\sqrt{\left(1-\frac{2Mr(d)}{\rho^2}+\Omega_h\left(-\frac{\Sigma}{\rho^2}\sin^2{\theta}\Omega_h+\frac{4Mar(d)\sin^2{\theta}}{\rho^2}\right)\right)}\right)}}$ and from equation (\ref{r}), $r(d)=\frac{(r_++r_-)}{2}{\mathcal{W}\left(\frac{e^\frac{2(r_++r_-)\ln{(r_+-r_-)}+r_+-r_-+4d}{2(r_++r_-)}}{2(r_++r_-)}\right)}+{\frac{1}{4}(3r_++r_-)}$.

\par
\begin{figure}[ht!]
     \includegraphics[width=7cm]{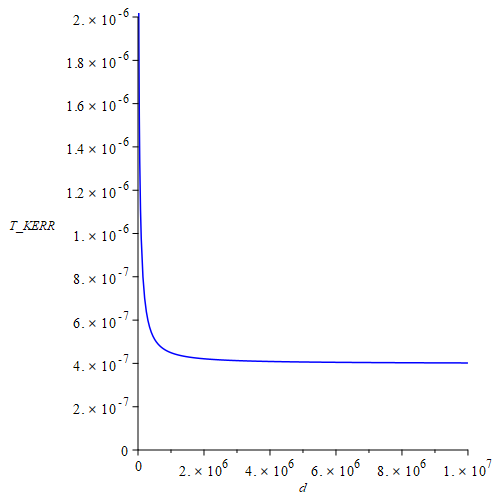}
     \caption{Here, we have plotted $T_{{loc}}$ as a function of the proper distance $d$ for the Kerr black hole. The parameters chosen for this plot are $M=10^5$ and $a=0.5$. With these parameter values, the corresponding Hawking temperature is $T_{\text{H}}=3.9788\times10^{-7}$. We have taken G=c=$k_B$=1.}
     \label{fig:KERR}
 \end{figure}

 Followed by equation (\ref{HawSpectum}), it is straightforward to see that the factor 
$\zeta_{KERR}(d)/2\pi$ essentially plays the role of the local temperature. Then, equation (\ref{KNFL}) can be recast as $\delta \mathsf{E}=T_{loc}\frac{\delta A}{4}$. This certainly establishes the fact that, from the perspective of a local observer at a finite distance, equation (\ref{KNFL}) for Kerr BH can be interpreted as the first law of black hole thermodynamics.\\
We have made a plot of the local temperature, $T_{loc}$, as a function of proper distance (see \ref{fig:KERR}). It is evident from the plot that for a massive Kerr black hole, the local temperature approaches and closely aligns with the Hawking temperature at larger distances.

\par Indeed, the similarity in the structure of the first law for different types of black holes, such as the Kerr and Reissner-Nordström black holes, indicates a universal framework for expressing the first law of black hole thermodynamics. This framework allows us to establish a consistent relationship between changes in black hole energy, and horizon area. This universal framework provides a powerful tool for studying black hole dynamics and understanding the interplay between gravity, thermodynamics, and the properties of spacetime.

\subsection{BTZ BH}
In this section, we will extend our previous study to the 2+1 dimensional BTZ black hole. While the BTZ black hole may not have as much astrophysical significance as its higher-dimensional counterparts, it has played a crucial role in theoretical physics for several reasons. Firstly, the BTZ black hole is an important object in the context of the AdS/CFT correspondence\cite{Maldacena:1997re}\cite{ammon_erdmenger_2015}\cite{Maloney_2010}. It serves as a simpler yet non-trivial example of a black hole in Anti-de Sitter (AdS) space, allowing researchers to explore various aspects of this correspondence and gain insights into the connection between gravity and quantum field theories. Secondly, the BTZ black hole has been used as a valuable tool in investigating the information paradox\cite{Yu_2022}. As a toy model, it provides a simplified setting for studying information loss and the potential resolutions of the paradox. Researchers have employed BTZ black holes in various scenarios and thought experiments, shedding light on the fundamental nature of black hole evaporation and information preservation. Lastly, the BTZ black hole offers a mathematical simplicity that facilitates its analysis compared to higher-dimensional black holes described by general relativity. This advantage allows for more tractable calculations and a deeper understanding of black hole thermodynamics and geometric properties. Overall, the BTZ black hole serves as an important theoretical laboratory for exploring fundamental concepts in gravity, gauge theories, and the interplay between quantum mechanics and gravity.
\par The metric of the $2+1$ dimensional BTZ black hole \cite{PhysRevLett.69.1849},\cite{SCarlip_1995} is given by, 
\begin{equation}
\label{btz}
 ds^2=-\mathcal{N}^2dt^2+\frac{1}{\mathcal{N}^2}dr^2+r^2(d\phi+\mathcal{N}^\phi dt)^2
\end{equation}
Where $\mathcal{N}^2=\left(-M+\frac{r^2}{\ell^2}+\frac{J^2}{4r^2}\right)$ and $\mathcal{N}^\phi=-\frac{J}{2r^2}$. The parameter $\Lambda=\frac{1}{\ell^2}$, represents the cosmological constant. The spacetime given by (\ref{btz}) exhibits coordinate singularities at $r=r_\pm$, which corresponds to the horizon of the BTZ black hole,
\begin{equation}
    r_\pm=\sqrt{\frac{M\ell^2}{2}\left(1\pm\left[1-\frac{J^2}{M^2 \ell^2}\right]^\frac{1}{2}\right)}
\end{equation}
The killing vector field to this spacetime which vanishes at the horizon, is given by 
\begin{equation}
    k^\alpha=\partial_t^\alpha-\mathcal{N}^\phi_h\partial_\phi^\alpha
\end{equation}
Where $\mathcal{N}^\phi_h=-\frac{J}{2r_+^2}$.
The norm of this killing vector field is given by,
\begin{equation}
\label{norm}
    \lvert\lvert k^\alpha \rvert\rvert=\sqrt{-\mathcal{N}^2+(\mathcal{N}^\phi)^2 r^2+(\mathcal{N}^\phi_h)^2 r^2+2\mathcal{N}\mathcal{N}^\phi_h r^2}
\end{equation}
The surface gravity for this BH is given by,
\begin{equation}
    \kappa=\frac{r_+^2-r_-^2}{\ell^2r_+}
\end{equation}
Now we can calculate the proper distance($d$) between the observer and the BH along a curve whose tangent vector field is $\partial_r^\alpha$. Then,
\begin{equation}
\begin{split}
d&=\int_{r_+}^r \frac{1}{\sqrt{\mathcal{N}^2}}dr\\
&=\int_{r_+}^r \frac{r\ell}{\sqrt{(r^2-r_+^2)(r^2-r_-^2)}}dr\\
&=\ell \log\left[1+\sqrt{\frac{r^2-r_+^2}{r^2-r_-^2}}\right]
\end{split}
\end{equation}
By rearranging this equation, we can solve for $r$ and express it in terms of the proper distance and the horizon radius. Subsequently, we obtain:
\begin{equation}
    r=\sqrt{\frac{r_+^2-(e^{\frac{d}{\ell}}-1)^2r_-^2}{1-(e^{\frac{d}{\ell}}-1)^2}}
\end{equation}
The result is elegant, straightforward, and practical. Unlike before, we do not rely on any approximations to express '$r$' as a function of the proper distance, and we also avoid the use of complex functions such as Lambert W. This simplification allows us to easily express equation (\ref{norm}) as a function of $d$, ensuring that the proportionality factor in equation (\ref{firstl}) is solely dependent on the proper distance. Consequently, we can express equation (\ref{quasi1st}) as follows:
\begin{equation}
\label{btzf}
    \delta \mathsf E=\frac{\zeta_{BTZ}(d)}{8\pi}\delta A
\end{equation}
Where $\zeta_{BTZ}(d)=\frac{\kappa}{ \lvert\lvert k^\alpha \rvert\rvert}$, and $\lvert\lvert k^\alpha \rvert\rvert$ is a function of of $d$.
\par Here, $T_{{loc}}=\frac{\zeta_{{BTZ}}(d)}{2\pi}$ represents the local temperature observed by a nearby observer at a finite distance from the BTZ black hole.

\section{Discussion}
In this paper, we made a thorough examination of the quasilocal first law. This law states that, for an observer close to a black hole horizon, the first law of black hole mechanics reduces to a form which involves variation of energy and area only. More precisely, the changes in the black hole's energy or mass are determined by the variation of its area of the horizon times a local temperature, regardless of the other charges, such as angular momentum, electric charge, etc. that the black hole might possess. 

We have investigated some implications of this law for an observer located at a finite proper distance from the black hole horizon. Our analysis is based on the specific cases of Reissner-Nordström (RN), Kerr, and BTZ black holes. Our findings demonstrate that the change in the black hole's energy is directly proportional to the variation of its area times a local temperature which solely depends on the proper distance from the black hole's event horizon.

To strengthen the claim of the local first law of black hole thermodynamics, we initially provide a brief discussion of the local temperature and subsequently demonstrate how the first law arises for each black hole of our study along with the corresponding local temperatures.

A careful observation of the local version of the black hole's first law reveals that it involves quantities that are defined on the horizon (area) and at the location of the observer (energy and temperature).

We have two different coordinate systems, the Eddington-Finkelstein coordinate system for RN black hole and the Boyer-Lindquist coordinate system for Kerr black hole, in order to demonstrate that our result is not a coordinate artifact.

Moreover, the local first law demonstrates how the original first law of black hole mechanics scales with observer's location. More specifically, we get a law of scaling such that the relation $\beta_H E_{\infty} = \beta_{\text{local}} E_{\text{local}}$ remains scale invariant. Upon examining the local energy, it is evident that other hairs, such as charge, angular momentum, etc., get absorbed into local energy $E_{\rm local}$, resulting in a simplified first law, while preserving the spectral distribution structure which exhibits a Planckian behavior. % In particular, the original first law and the Hawking temperature is recovered when the observer is located at asymptotic infinity. 
To our knowledge, this scaling behaviour of the first law of black hole has not been pointed out in any earlier literature and is a new result. We plan to study this scaling property in our future work.
\nocite{*}
\bibliographystyle{ieeetr}
\bibliography{main}% Produces the bibliography via BibTeX.
\end{document}